# Morphotectonic Study of the Greater Antilles

**Mario O. Cotilla Rodríguez\*, Diego Córdoba Barba, and Diana Núñez Escribano**

*Departamento de Física de la Tierra, Astronomía y Astrofísica I (Geofísica y Meteorología), Facultad de Ciencias Físicas, Universidad Complutense de Madrid, Ciudad Universitaria, 28040 Madrid, Spain*
*\*e-mail: macot@ucm.es*

**Abstract**—The first morphotectonic model of the Greater Antilles is presented. The model is adjusted to the current dynamics between the Caribbean and North American plates. It is mainly elaborated by Rantsman's methodology. We determined 2 megablocks, 7 macroblocks, 42 mesoblocks, 653 microblocks and 1264 nano-blocks. They constitute a set of active blocks under rotation, uplifting and tilting movements. A total of 11 active knots of faults and 8 cells are the main articulation areas. The largest seismogenetic structures in the Northern Caribbean are an array of the active fault segments. The majority of them are in the Caribbean-North American Plate Boundary Zone, the Hispaniola has the most complex neotectonic structure–associated with the central axis of the morphotectonic deformations in the region.

## INTRODUCTION

The Northern Caribbean (Fig. 1) is an example of the diversity of seismotectonic process that mainly correspond to regional dynamics found within a framework of different category plate interactions. In it there are four islands: Cuba (C), Hispaniola (H), Jamaica (J) and Puerto Rico (PR). All they are known as the Greater Antilles Arc ($\sim$209.400 km$^2$). H includes two countries (Haití and the Dominican Republic).

The Caribbean Sea has an area of $\sim$2800000 km$^2$ and its northern part is distinguished by the presence of three troughs (T) (Cayman, T and Muertos). In the eastern extreme of the large Cayman trough the Oriente trough is developed. This entire region has been investigated of fragmentary form and attending to different interests, the majority of them far from of their citizens. Nevertheless, these investigations allowed getting a relatively clear picture on the main tectonic and seismic characteristics.

Seismotectonic process (ST) is a scientific specialty of the great importance for the planning large-scale industrial projects, tourist complexes and for the preservation of human life. Unfortunately, to date no unified ST methodology exists. It is explained by several reasons but we only indicate four: (1) the necessity to propose solutions and answers from the consequences of geological processes; (2) the great diversity of the geodynamic conditions in which the earthquakes take place; (3) the insufficient and heterogeneous seismological devices and historic data; (4) the high cost of the geologic, tectonic and geophysical data.

A ST study does not limit investigation to the narrow geographical confines of a country. Then, it is necessary to look up to a ST Province, to clarify its position in the hierarchical structure. The elaboration of a seismotectonic map (SM) is not limited to apply the same methodology in the different seismic zones. It does not mean that it is considered incorrect to use relationships and formulas deduced for other regions to establish a maximum magnitude depending on the geometrical characteristics of the fault. However, the results should then be critically evaluated.

The main objective of our work is to expose a homogeneous morphotectonic model for the aforementioned islands, like an initial step in the preparation of a SM. This model is based on a same methodology, resulting in coherence within the delimited units.

## MAIN GEOLOGIC AND TECTONIC CHARACTERISTICS

For the preparation of this part the following works were used [3, 7–13, 23, 26, 31–38, 40, 48–50, 54, 55, 57, 61–63, 65, 66, 68, 71]. The most relevant data are two: (1) the relative displacement of the Caribbean and North American plates controls the tectonic regime; (2) the northern and southern borders of the Caribbean plate show quite different characteristics.

The Northern Caribbean arc was developed during the Lower Cretaceous and the magmatic activity mostly finished in the Lower Eocene as a consequence of the direct interaction with the Bahamas Platform. With the NE movement of the Caribbean block in the Eocene the subduction process produced the volcanic arcs of the Cayman Cordillera and the Sierra Maestra Mountains. Then, a strike-slip fault system (Cayman) started. The convergence finished and the ST of the Caribbean region was mainly determined by the interaction of the North American, South American, Cocos, Nazca and Caribbean plates.

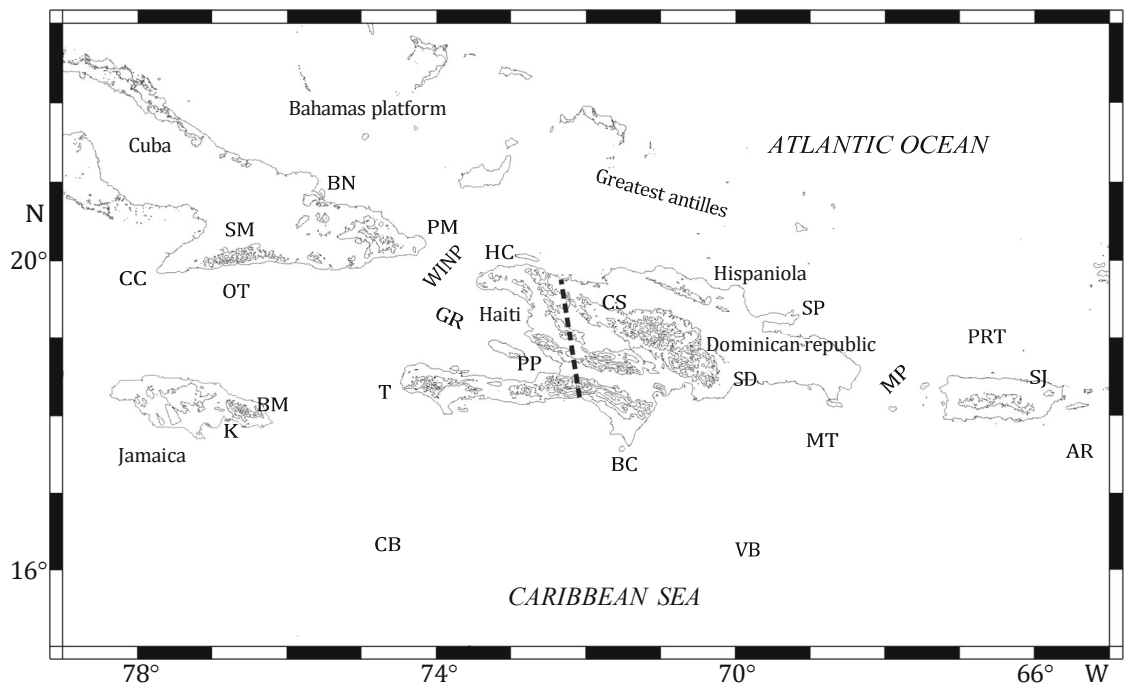

**Fig. 1.** The Septentrional Caribbean. Central-Eastern Greater Antilles Arc with: basins: CB–Colombia, VB–Venezuela; cities: K—Kingston, PP—Puerto Principe, SD—Santo Domingo, SJ—San Juan; localities: BC—Beata Cape, BN—Bahía de Nipe, CC—Cabo Cruz, HC—Haitian Cape, PM—Punta de Maisí, SP—Samaná Peninsula: T—Tiburón; mountains: BM—Blue, CS—Central System, SM—Sierra Maestra; passages: MP—Mona, WINP—Wins; ridges: Aves, Beata; troughs: MT—Muertos, OT—Oriente, PRT—Puerto Rico.

The Cayman strike-slip fault system is divided into 2 branches: (1) northern branch (from the Cayman trough to the spreading centre and then to the PR trough); (2) southern branch (from Central America to Haití) (Fig. 2a). The northern branch is active. The western part of the southern branch (Walton-Plantain Garden-Enriquillo fault zone (EPGFZ)) is clearly active and runs from J up to the Muertos trough. Within the North American-Caribbean Plate Boundary Zone (PBZ) 2 microplates were defined (Gonave and H-PR) (Figs. 1 and 2a). A continuous, northern strike-slip fault system (the Oriente fault zone, Cibao Valley faults, Northern H-Septentrional fault zone) bordering both the Gonave and the H-PR microplates runs from the northern shore of Haití to the PR island slope and the PR trough. Two other smaller microplates (El Seibo (eastern H) and PR) are also found here. There is a great difference in spread (17 and 2 mm/yr) compared to the North American plate's oblique NE-SW subduction (PR trough) and the Caribbean plate in the Muertos trough subduction zone, respectively.

The western area of H is characterised by microplate crustal convergence and strike-slip faulting accompanied by Quaternary tectonic uplift. But the eastern area is composed of the subducted Atlantic and Caribbean ocean floor with small or very small speed and a clear Quaternary tectonic uplift. These different geodynamic mechanisms favor lateral, vertical and relative-turn movements within the H megablock. The fracturing, seismicity and focal mechanisms may be explained inside of a PBZ of ~200 km wide.

The Beata Ridge is another oceanic structure of high interest to our morphotectonic model. It extends 400 km from the South of Beata Cape up to northern H. This strcuture divides the Caribbean into the Columbia and Venezuela basins. In the West, the ridge is bound by a steep escarpment with a regional slope of 15°–25° which rises 2500 m above the Columbia abyssal plain. By the East, the ridge drops down to the centre of the Venezuela basin in different steps. The Beata Ridge is an oceanic plateau whose edges have been reactivated by differential motion between the two aforementioned structures. The Gonave, Northern H and PR microplates are attached to the Venezuela microplate.

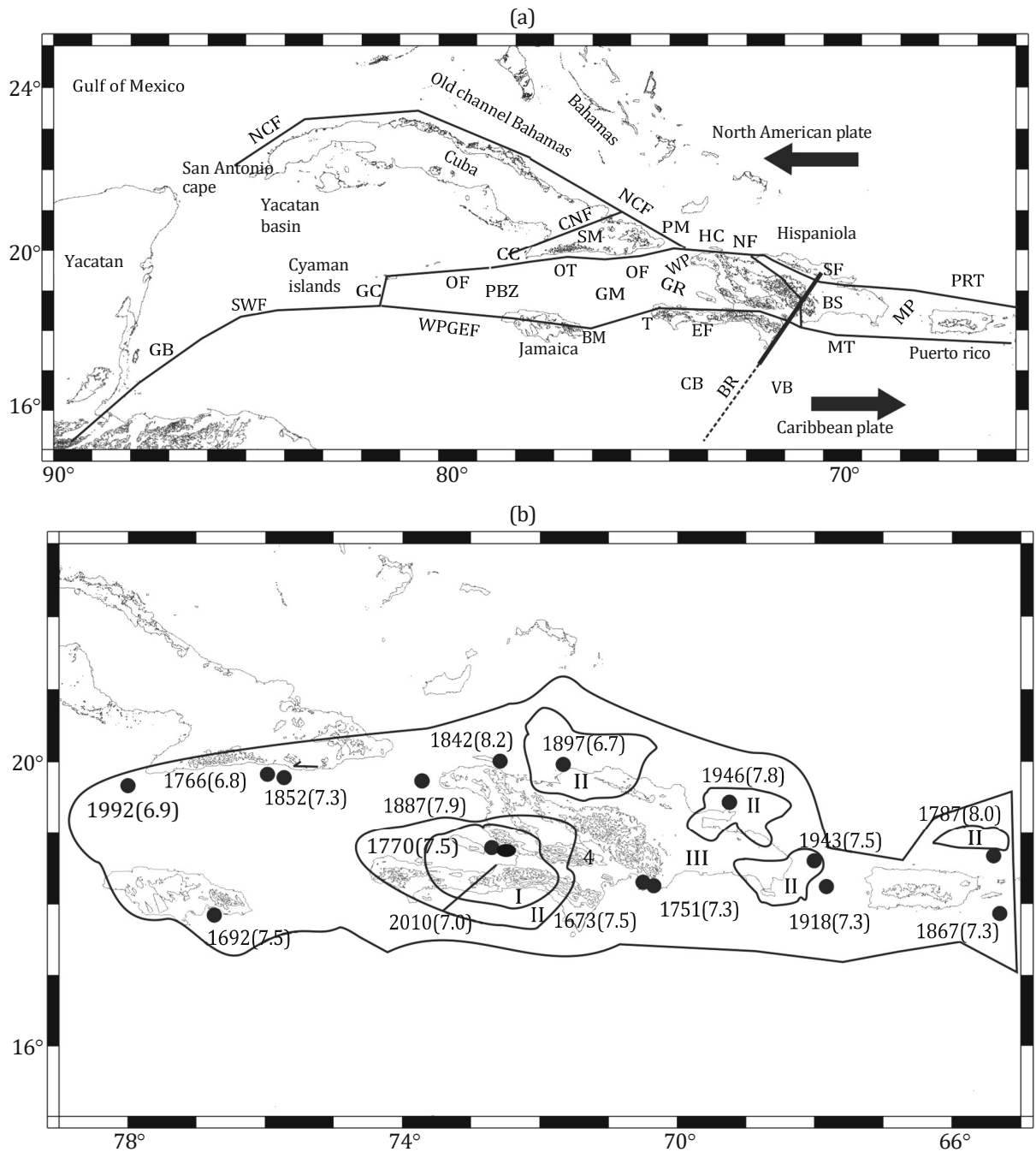

**Fig. 2.** Scheme of the tectonic Caribbean-North American area (a) and selection of strong earthquakes with epicenter density (2008–2012) (b). Arbitrary notes for panel (a): basins: CB—Colombia, GB—Guatemala, VB—Venezuela; black lines denote; faults: CNF—Cauto-Nipe, EF—Enriquillo, NCF—Nortecubana, NF—Northern Hispaniola, OF—Oriente, SF—Septentrional, SWF—Swan-Walton; heavy black arrows, directions of plate movements; localities: CC—Cabo Cruz, HC—Haitian Cape, PM—Punta de Maisí; mountains: BM—Blue Mountain, SM—Sierra Maestra; other structures: BS—Beata escarpment, BR—Beata Ridge, GC—Mid-Cayman Spreading Centre, GM—Gonave microplate, GR—Gonave Rise; passages: MP—Mona, WP—Wins; troughs: MT—Muertos, OT—Oriente, PRT—Puerto Rico. Arbitrary notes for panel (a): black circles denote earthquake epicenters, with the year of occurrence and magnitudes (in brackets) indicated; epicenter density: 3rd iteration: I, 0.1–0.5; II, 0.4–0.2; III, <0.2.

The two deepest ocean troughs (Fig. 1) in the Caribbean are the Oriente (∼–7500 m) and the PR (∼–8800 m). The second trough (PR): (1) is deeper and larger (PR = ∼1500 km, Oriente = ∼350 km); (2) has deeper earthquake foci and a larger epicenter density; (3) produces tsunamis. These troughs belong to two macroblocks of the Northern Caribbean plate, but their geodynamic context is different. Another significant regional difference is the neotectonic gradient of relief. The Oriente trough has an important and direct relationship with the adjacent relief of the Eastern C macroblock (hmax = 1974 m). The total altitude difference is approximately 10 km while that is not the case in the PR segment. These data indicates that lateral displacement of the Caribbean and North American plates have important and different movements (vertical, direct and reverse). Also, there are rotation of blocks and microplates. For example, PR and its surroundings are more adjusted to block rotations, while in Eastern C vertical movements prevail.

**Table 1.** The main earthquakes of the Northern Caribbean

| Island | Date | Magnitude | Fatalities |
|---|---|---|---|
| Cuba | 1852.08.20 | 7.3 | 5 |
| | 1992.05.25 | 6.9 | – |
| | 1766.06.11 | 6.8 | 8 |
| | 1932.02.03 | 6.7 | 7 |
| | 1947.08.07 | 6.7 | – |
| Jamaica | 1692.06.07 | 7.5 | 2000 |
| | 1943.07.28 | 7.5 | Unknown |
| | 1907.01.14 | 6.5 | 1600 |
| | 1957.03.02 | 6.5 | Unknown |
| | 1914.08.03 | 6.0 | Unknown |
| Puerto Rico | 1670.08.15 | 8.0 | Unknown |
| | 1787.05.02 | 8.0 | Several |
| | 1867.11.18 | 7.3 | Unknown |
| | 1918.10.11 | 7.3 | 140 |
| | 1974.10.08 | 7.1 | – |
| Hispaniola | **1842.07.05** | **8.2** | **~500** |
| | 1615.09.08 | 8.0 | 150 |
| | 1887.09.23 | 7.9 | ~220 |
| | 1946.08.04 | 7.8 | 10 |
| | 1673.05.09 | 7.5 | ~500 |
| | 1684 | 7.5 | Unknown |
| | 1770.06.04 | 7.5 | 100 |
| | 1943.07.29 | 7.5 | ~20 |
| | 1751.10.18 | 7.3 | ~2000 |
| | 1948.04.21 | 7.3 | Unknown |
| | 1953.05.31 | 7.0 | Unknown |
| | 2010.01.12 | 7.0 | ~31000 |
| | 1691 | 6.7 | Unknown |
| | 1987.12.29 | 6.7 | Unknown |
| | 1701.11.09 | 6.6 | Unknown |
| | 1751.11.27 | 6.6 | 5000 |
| | 1910.05.11 | 6.5 | Unknown |
| | 1911.09.11 | 6.5 | ~100 |
| | 1860.08.09 | 6.3 | Unknown |
| | | ∑34 events | **~40000** |

**Table 2.** Data of strong earthquakes

| Island | Mmax | Main Events | Total period of occurrence (year) | Fatalities |
|---|---|---|---|---|
| Cuba | 7.3 | 5 | 226 | 20 |
| Hispaniola | **8.2** | **19** | **395** | **~35000** |
| Jamaica | 7.5 | 5 | 265 | ~3600 |
| Puerto Rico | 8.0 | 5 | 304 | ~300 |
| | ∑ | 34 | – | ~40000 |

The third trough in the area is called Muertos. It is an east-west striking oceanic structure of 5 km depth that defines the southern limit of PR-Virgin Islands. A N dipping zone of earthquakes to a depth of 150 km and the accretionary prisms along the lower slope S of southeastern H and southwestern PR are in concordance with the overriding of the Caribbean plate by southwestern PR-Virgin Islands along the aforementioned depression.

From a neotectonic point of view, C belongs to the southern part of the North American plate, and the rest of the previously mentioned islands are in the Caribbean plate. However, all they are spatially near and under the dynamic influence of the 5 plates above mentioned.

## SHORT REVIEW OF THE SEISMICITY

To develop this epigraph we used some works: [5, 13–16, 18, 27, 46, 47, 51–53, 58, 59, 62, 65, 67, 69, 71, 73]. Tables 1 and 2 reflect the most important earthquakes for the islands. The maximum magnitude value is 8.2 in H. The total quantity of fatalities that we estimated is ~40000.

In the Northern Caribbean the most intense seismicity is located around restraining bends such as southern C and northern H (Fig. 2B). Three examples: (1) all Cuban earthquakes of the tables 1 and 2 were produced in the southeastern part; (2) the largest seismic activity (SA) is determined from H to PR. This segment has 33 strong events (8.2–8.0 = 4, 7.9–7.0 = 16, 6.9–6.3 = 13); (3) the foci distribution of the August (8–24) – September (25) – October (4) of 1946 H earthquakes series (Ms = 7.3, Ms = 6.2, Ms = 5.4 and Ms = 6.1) has a seismogenic layer depth of 5–35 km in 2 long bands of 250 × 75 km, oriented approximately WNW. It is parallel to the Septentrional fault zone and South Samaná fault.

The level of historical seismicity in the Caribbean is to date higher than the instrumental period. Cotilla [15] interpreted it as over estimation and mistakes. The majority of the countries (C, J, Dominican Republic and PR) has a permanent seismic network. PR has the best array and equipments. All the Caribbean islands have interplate (or plate edge) seismicity, while C has also intraplate seismicity (Table 1).

The great majority of earthquakes in the Caribbean are associated with the PBZ. The largest area of Caribbean is aseismic. The rate of earthquakes at the limit of the Cocos-Nazca-Caribbean plates in relation to the North American-Caribbean plates is 130/15; while with regard to the North American-Caribbean-South American plates, it is 51/2. These data allow a rough estimation of microplate movements inside the Caribbean plate.

A selection of focal mechanisms of the Northern Caribbean is represented in figure 8.5 of Cotilla [15]. It was revised by Núñez et al. [55] and it permitted to understand the differences of movements and the tendency of the blocks inside the PBZ, mainly in H and PR.

We show a relationship of 29 tsunamis in the Northern Caribbean (H = 13, J= 8, PR = 6 and C = 2) (Table 3). A rate of 0.09 events/year is a low value in comparison with the Caribbean Pacific region (1.3 events/year). The highest value corresponds to H and the total fatalities for such events in this region is ~10000. Taking into account the occurred strong earthquakes and tsunamis, we can state that the aforementioned faults are active and of the first order.

## MORPHOTECTONIC ANALYSIS

### Methodology

Ranstman's methodology was applied and explained in the following works [2, 4, 6, 17, 21, 24, 39, 43, 60]. The methodology can be summarized as follows: (1) delineation of Territorial Units (TU) of 3 types (superficial, linear and linear intersections); (2) the superficial TU has a developed hierarchy (megablock, macroblock, mesoblock, block, microblock and nanoblock); (3) the linear TU is the limit of the superficial one and it has length and strike. This element does not always coincide with known faults; (4) the intersections of the linear TU are called knots. A knot is usually the most active zone and represented by a circular figure. To obtain these 3 elements (multi-scale) topographic maps, aerial photos, satellite images and photos are used. Then, it may possible to develop a set of special maps and schemes of the relief. These data are compared and complemented by the geophysical, geologic, geomorphologic and tectonics results.

As an alternative to current tectonic research we have taken some elements as lineaments and interception of lineaments (knots) using Remote Sensing methods Aduskin et al. [1]. The original idea is justified by the results of some specialists [42, 44, 45, 64, 70, 74] that other researchers also applied.

### Main Characteristics

The majority of the data used are from [13, 14, 17, 19, 21–25, 28–30, 41]. By taking advantage of the results of all the previous research methods and using Rantsman's methodology, a suitable framework was set up, potentially establishing the morphotectonic regionalisation of C, H, J, and PR. Some data of interest appear in Table 4 and information regarding the Main Superficial Waters Divide of First-Order (MDFO) in the islands are shown in Figs. 3 (a, b, c and d).

Cotilla and Córdoba [20] applied the Okubo and Aki [56] fractal methodology to southeastern Cuba. They also used the Wice et al. [72] technique of box-counting. As the results were very reliable we also determined the fractal dimension (D) of the region studied. Using strike analysis of the fractures and faults we achieved the following values of D: Eastern C = 1.73, H = 1.98, J = 1.85 and PR = 1.81. The larger values of D are associated with more complex fault geometry. From these values we concluded that the region is tectonically complicated and the highest level corresponds to H.

### Cuba Island

The area of Cuba (110 911 km$^2$) is higher than the total of the other islands (98 422 km$^2$). However, it is different in respect to the longitude of the MDFO (C = 1260 km/3 islands = 2206 km). The figure of C is concave (SW-NE to NW-SE) as a consequence of the ancient convergence previously referred.

C belongs to the North American plate and it is a ST province in the Caribbean. We determined 3 neotectonic macroblocks (Western, Central-Eastern and Eastern) and 3 knots (NMG0, NMG1 and NMN1) in the Cuban megablock. The last macroblock is the most active. Figure 4 shows the location of these structures and their limits. The other TU are: 24 mesoblocks, 415 blocks, 542 microblocks and 617 nanoblocks. Also, the adjacent marine surroundings of C have a clear block structure. The MDFO of the fluvial network is quite regular from the Western macroblock to the Central-Eastern macroblock, but in the Eastern one exists 2 branches. Such characteristic is related to a high level of neotectonic activity.

The limit between the Central-Eastern and Eastern macroblocks is a 2nd order fault (Cauto-Nipe). This fault has 2 seismoactive knots (NMGO and NMN1) at its ends (Fig. 4). They connect the Oriente fault with the Cauto-Nipe fault in the SW near the Cabo Cruz and Nortecubana fault with the Cauto-Nipe fault in the NE to the north of Bahía de Nipe, respectively. The first knot is associated with the earthquake of 25.05.1992 (Ms = 6.9). Using these elements, it has been possible to establish a geodynamic cell to Eastern C in order to justify the SA.

The current structure of Eastern C (~20000 km$^2$) includes extremely diverse areas, differing in layout,

**Table 3.** Tsunamis of the Northern Caribbean

| No. | Date | Site | Classification |
|---|---|---|---|
| 1 | 1690.04.16 | 17.5 N/61.5 W; U.S. Virgin Islands | S |
| 2 | 1692.06.07 | 17.8 N/76.7 W; Jamaica (Port Royal, Liganee (Kingston), Saint Ann's Bay) | S |
| 3 | 1751.10.18 | 18.5 N/70.7 W; Hispaniola (Dominican Republic Azua de Compostela, Santo Domingo, Santa Cruz, El Seíbo) | S |
| 4 | 1770.06.04 | Haiti (Golfo de la Gonave and Arcahaie) | S |
| 5 | **1842.05.07** | **19.7 N/72.8 W;** Haiti (Mole Saint-Nicolas. Haitian Cape, Port-de-Paix. Forte-Liberte), Dominican Republic (Santiago de los Caballeros, Santo Domingo, Northern coast of Hispaniola) | S |
| 6 | 1860.03.08 | 19.0 N/72.0 W; Hispaniola (Gulf de la Gonave, Les Cayes, Acquin, Anse-a-Veau) | S |
| 7 | 1867.11.18 | 18.0 N/65.5 W; Virgin Islands (St. Croix, St. Thomas) | S |
| 8 | 1887.09.23 | 19.7 N/74.4 W; Haiti (Mole Saint-Nicolas, Jeremie, Anse-d'Hainault, Point Tiburón) | S |
| 9 | 1907.01.14 | 18.1 N/76.7 W; Jamaica (Annotto Bay, Bluff Bay, Hope Bay, Orange Bay, Sheerness Bay, St. Ann's Bay, Kingston, Ocho Rios, Port Antonia, Port Maria) | S |
| 10 | 1918.10.11 | 18.5 N/67.5 W; Puerto Rico (Aguadilla, Bahía de Boquerón, Cayo Cardona, Guanica, Isabella, Isla Caja de Muertos, Isla Mona, Mayagüez, Playa Ponce, Puerto Arecibo, Punta Agujereada, Punta Borinquen, Punta Higüero, Río Culebrinas, Rio Grande, Río Grande de Loiz,), Dominican Republic (Santo Domingo (Río Ozama)) | S |
| 11 | 1918.10.24 | 18.5 N/67.5 W; Puerto Rico (Galveston, Mona Passage, Texas) | S |
| 12 | 1939.08.15 | 22.5 N/79.2 W; Cuba (Cayo Francés) | S |
| 13 | 1946.08.04 | 19.3 N/68.9 W; Dominican Republic (Cabo Samaná, Julia Molina, Matancitas), Haiti and Puerto Rico (San Juan) | S |
| 14 | 1946.08.08 | 19.5 N/69.5 W; Puerto Rico (Aguadilla, Mayagüez, San Juan) | S |
| 15 | 1989.11.01 | 19.0 N/68.8 W; Puerto Rico (Cabo Rojo, East of Nuevo Día) | S |
| 16 | 1688.03.01 | Jamaica (Port Royal) | P |
| 17 | 1751.09.15 | 18.5 N/70.7 W; Hispaniola (Haiti) | P |
| 18 | 1751.11.21 | 18.3 N/72.3 W; Haiti (Port-au-Prince) | P |
| 19 | 1769 | 18.5 N/72.3 W; Haiti (Port-au-Prince) | P |
| 20 | 1775.02.11 | 19.0 N/72.4 W; Hispaniola, Cuba | P |
| 21 | 1775.03 | 19.0 N/72.3 W; Hispaniola | P |
| 22 | 1775.12.18 | 19.2 N/70.3 W; Hispaniola, Cuba | P |
| 23 | 1780.10.03 | 18.1 N/78.1 W Jamaica (Savanna La Mar) | P |
| 24 | 1781.08.01 | 18.2 N/78.1 W; Jamaica (Montego Bay) | P |
| 25 | 1787.10.27 | 18.4 N/77.9 W; Jamaica (Montego Bay) | P |
| 26 | 1812.11.11 | 18.0 N/76.5 W; Jamaica (Annotto Bay) | P |
| 27 | 1881.08.12 | 19.9 N/76.8 W; Jamaica (Kingston) | P |
| 28 | 1931.10.01 | 21.5 N/80.0 W; Cuba (Las Villas, Playa Panchita, Rancho Veloz) | P |
| 29 | 1953.05.31 | 19.7 N/70.7 W; Dominican Republic (Puerto Plata) | P |
|  |  | ∑S = 15, ∑P = 14 |  |

P = Probable, S = Sure.

**Table 4.** Main characteristics of the Territorial Units

| Parameters | C | H | J | PR | ∑ |
|---|---|---|---|---|---|
| Area, km$^2$ | 110 922 | 77914 | 11424 | 9104 | 209364 |
| Megablocks | 1 | 1 | – | – | 2 |
| Macroblocks | 3 | 2 | 1 | 1 | 7 |
| Mesoblocks | 24 | 14 | 2 | 2 | 42 |
| Blocks | 415 | 209 | 11 | 18 | 653 |
| Microblocks | 542 | 401 | 29 | 31 | 1.003 |
| Nanoblocks | 617 | 527 | 65 | 55 | 1.264 |
| Close relief surfaces | 3.482 | 1.969 | 1.022 | 534 | 7.007 |
| Fluvial basins (northern/southern) | 81/112 | 166/187 | 22/36 | 22/36 | 291/371 |
| Fracture density | 97.3 | 174.2 | 98.5 | 87.8 | 457.8 |
| Fractal dimension | 1.73 | 1.98 | 1.85 | 1.81 | – |
| Longitude of the coast line, km | 5745 | 3498 | 800 | 501 | 10544 |
| Longitude of the MDFO | 1.260 | 1.700 | 300 | 206 | 3.266 |
| Main knots 3/4 orders | 10/13 | 6/5 | 7/9 | 5/7 | 28/34 |
| Main strike of MDFO | E–W | E–W–NW | E–W | E–W | – |
| Main watersheds of 1st order (MDFO) | 2 | 4 | 1 | 1 | 8 |
| Maximum altitude, m | 1974 | 3175 | 2230 | 1338 | – |
| Maximum order of the main river | 7 | 7 | 7 | 6 | – |
| Maximum magnitude of earthquake | 7.3 | **8.2** | 7.5 | 8.0 | – |
| Maximum uplifting value, m | 1138 | 2235 | 1817 | 983 | – |
| River valleys (V/U) | 2.500/1.755 | 3.257/1.956 | 110/50 | 182/166 | 6. 049/3.927 |
| Sinuosity coefficient of the MDFO | 0.70 | 0.96–0.64 | 0.69 | 0.59 | – |
| Strong earthquakes | 5 | 19 | 5 | 5 | 34 |
| Tectonic localization | NP | PBZ–CP | PBZ–CP | PBZ–CP | – |
| Tsunamis | 2 | **13** | 8 | 6 | 29 |
| Uplifting areas 2–3/2–5 orders | 1.074/1.501 | 200/360 | 57/74 | 20/210 | 1.351/2.145 |

C—Cuba, H—Hispaniola, J—Jamaica, PR—Puerto Rico; CP—Caribbean plate; MDFO—Main Superficial Divided of First Order; NP—North American plate; PBZ—Plate Boundary Zone.

morphology and historical development. It began in the Late Eocene, on a mixed basement and, in general, on crusts of various thicknesses and types, ranging from sub-continental to sub-oceanic. The evolution of this structure was associated with, and considerably influenced by, deep-water troughs such as the Yucatan Basin in the SW, the Old Bahamas Channel in the NE and the Oriente trough in the S. Some of its geomorphologic characteristics are quite similar to Western H.

In C there are 12 active faults. The largest level of SA is associated with the Oriente fault (of 1st order). It is located in the SE border of the megablock. But in other zones of the island the earthquakes are related with other faults (i.e.: Nortecubana (28.02.1914, M = 6.2) and Guane (23.02.1880, M = 6.2), both of 2nd order) and knots (i.e.: Torriente-Jagüey Grande, N-TC (16.12.1982, M = 5.0)) (Fig. 4). No tsunamis are associated to the Eastern C macroblock. In 226 years of strong SA only 20 fatalities have occurred (Tables 1 and 2). This value is the lowest in the Caribbean region.

### Hispaniola Island

H is a megablock (Fig. 5) within the PBZ and bounded by the following structures: (1) in the west, the islands of C and J, the Oriente trough, the Gonave rise, the Navasa trough and the Pedro escarpment; (2) in the north, the Bahamas Platform, the North American plate and the Northern H fault zone-PR trough; (3) in the S, the Caribbean plate, the Muertos trough, the Colombian and Venezuelan basins and the Beata Ridge; (4) in the E the island of PR and the Mona Passage. Moreover, two macroblocks have been identified (Western and Eastern). Between the macroblocks is a transverse NE alignment, which we called Beata. It stands over the Beata Cape, Azua, Monseñor Miguel, Maimón, Cotui and Scottish Bay zones. It is considered a first morphostructural step of the megablock. Within the macroblocks there are 5 mesoblocks. The subdivisions of the mesoblocks signal the existence of 209 blocks, 401 microblocks and 527 nanoblocks. They are associated with 11 main active knots. There is also a block structure in the marine area, but it is quite different in the north with respect to the S. All these structures correspond to the above-mentioned Gonave microplate.

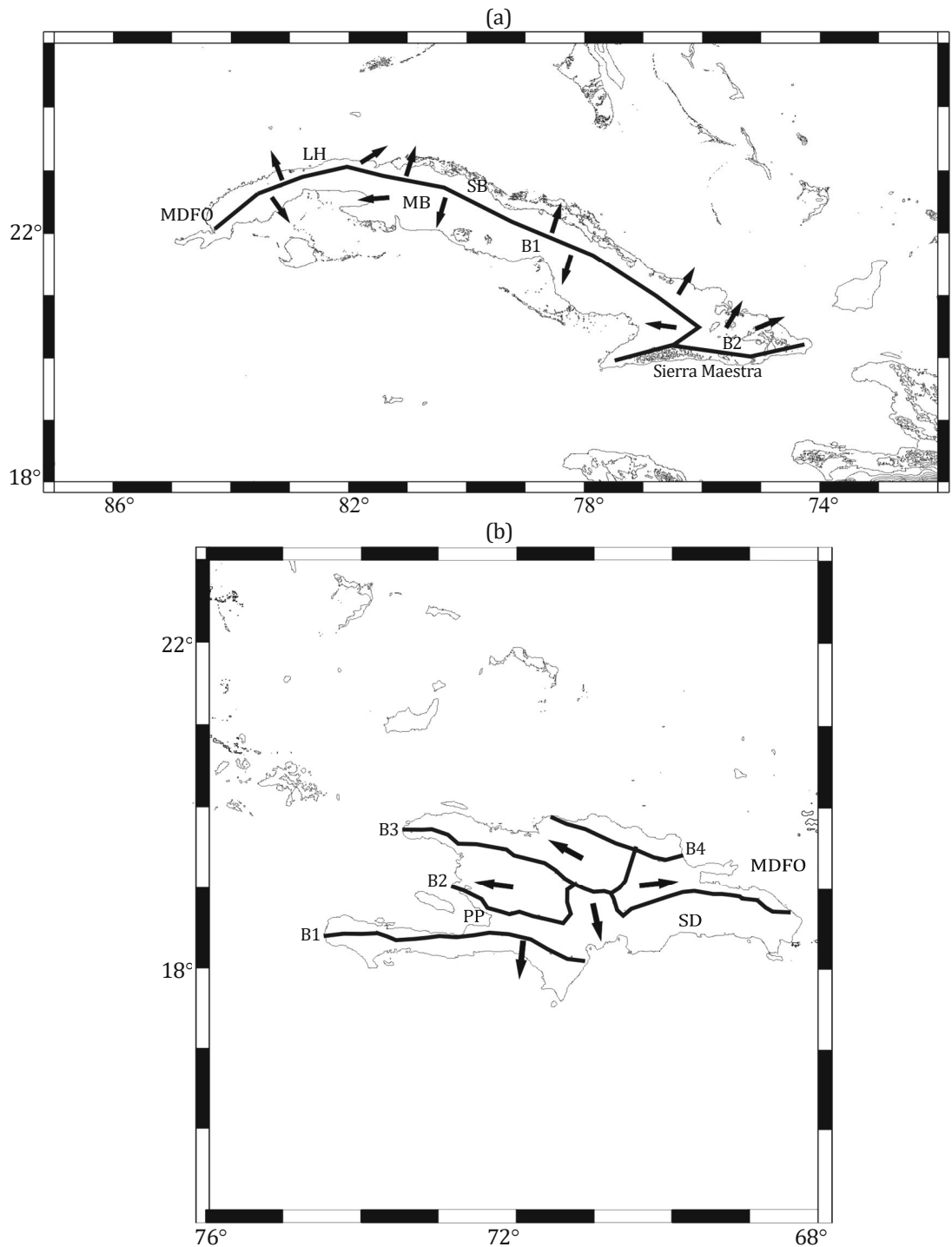

**Fig. 3.** The main first-order division of the islands: (a) Cuba, (b) Hispaniola, (c) Jamaica, (d) Puerto Rico branches (B1, B2); cities: Kingston (Jamaica), LH—La Habana (Cuba), PP—Puerto Principe (Haiti), SD—Santo Domingo (Rep. Dominicana), SJ—San Juan (Puerto Rico); heavy black arrow—sense of fluvial drainage; main basins: MB—Meridional, SB—Septentrional; MDFO—Main Superficial Water Divide of First Order, represented by a heavy black line. Branches B1, B2, B3 and B4 are used in Dominican Republic.

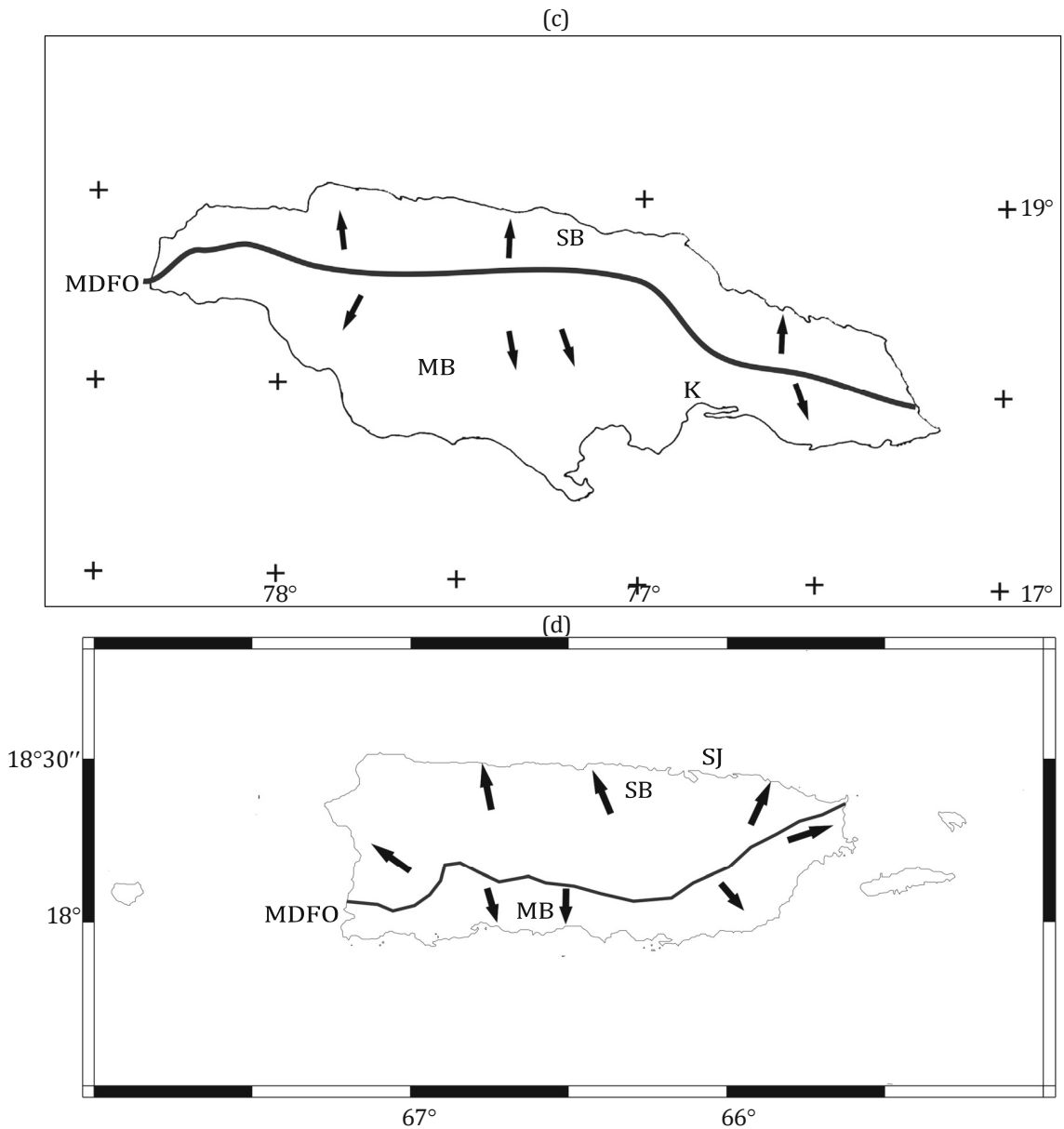

**Fig. 3.** (Contd.)

The MDFO of the fluvial network is a multi-parallel system with a dominant bearing of WNW to NW, similar to that of Eastern C (Fig. 3b). This fluvial system consists of 4 branches with a total length of 1700 km. This value is the largest of all islands. The branch lines are structurally related to recent forms in the relief.

The SA of H shows that in 395 years there have been 19 strong earthquakes with ~35000 fatalities and 13 tsunamis (Tables 1, 2 and 3). We determined that: (1) the main activity has been concentrated on the first- and second-rank lineaments; (2) some significant epicenters are located in the vicinity of the lineament intersections. This may be due to tension or compression in a restraining bend zone and by the forcing and/or pushing of macroblocks northeastwards.

### Jamaica Island

J shows that there is recognized control and neotectonic influence of the active regional Swan-Bartlett-Cayman (or Swan-Oriente) fault system (Fig. 6). A 2nd order morphostructure, it emerged from and is active in the PBZ of the Northern Caribbean. Two 3rd order morphostructures (Western and Eastern) were determined, the most active being the easternmost one. Additionally, 11 blocks, 29 microblocks and 65 nanoblocks were identified. This set forms a heterogeneous network of lineaments and 11 knots. Knots N1 (Montego) and N11 (Kingston) are the most active. There is a MDFO with an E-W strike (Fig. 3c), where major close surfaces are cut by transverse fractures. The greatest value of the neotectonic movements is in the Blue Mountains. It is a strong positive structure ~260 m higher than Sierra Maestra Mountains of C.

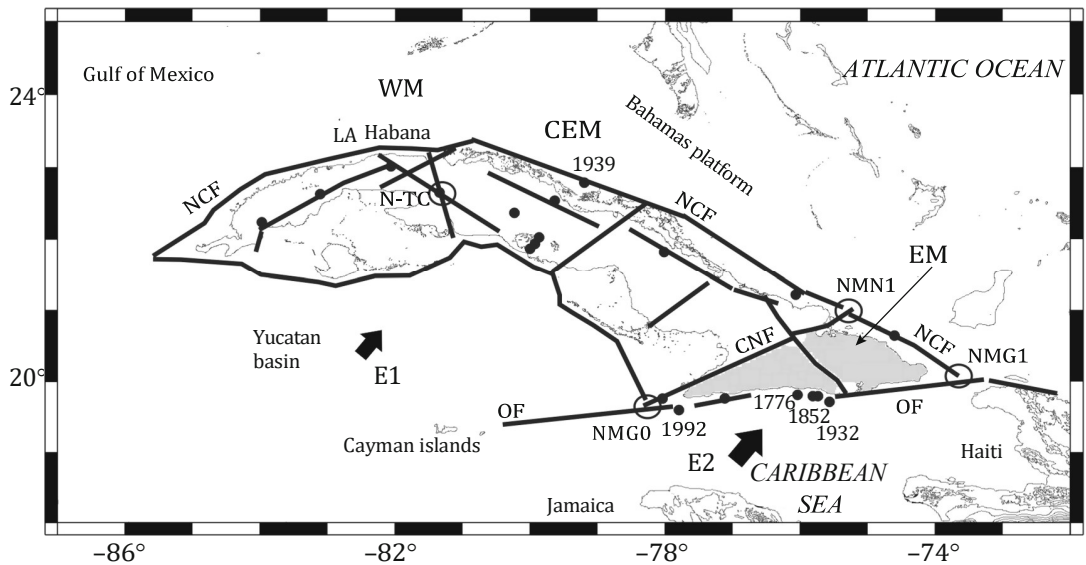

**Fig. 4.** Morphotectonic map of Cuba. Arbitrary notes: black circles denote epicenters; black lines, alignments of faults; empty circles, knots; faults: CNF—Cauto-Nipe, NCF—Nortecubana, OF—Oriente; knots: NMG0—Cabo Cruz, NMG1—Punta de Maisí, NMN1—Bahía de Nipe, N-TC—Torriente; macroblocks: CEM—Central-Eastern, EM—Eastern, WM—Western; digits near an epicenter mean occurrence year.

The contemporary faulting is more important in the marine parts of the N and S zones where the strongest earthquakes are generated. However, the relationship of this element to the disruptive structures, which were inherited from, modified or activated by the emerging parts is different. The SA is explained by its space-time temporary location in a transpression area of the PBZ and where 6 relevant seismogenetic zones exist. The larger geological hazard is framed by these elements. Six strong earthquakes occurred in 265 years, with numerous fatalities ~3600 (Tables 1 and 2). Also, J has associated 8 tsunamis (Table 3). These last numbers are quite important in order to understand the morphotectonic differences respect to C.

*Puerto Rico Island*

PR is a smaller sub aerial exposed part of the Greater Antilles Arc (Fig. 7). It is an emergent and tectonically active macroblock at the NE edge of the Caribbean-North American PBZ. The macroblock is asymmetric from the morphotectonic point of view and it is composed by 2 mesoblocks (Northern and Southern). The Northern mesoblock is larger and tectonically more active. Both structures include a total of 18 blocks, 31 microblocks and 55 nanoblocks. It was delimited 10 major lineaments and 83 lineament intersections, 12 of which are the main intersections (or knots). Such intersections are the most tectonically active and indicate fault segmentation, block rotation and low SA.

Seismic data indicates that 5 strong earthquakes were produced in 304 years with 6 tsunamis and ~300 fatalities (Tables 1, 2 and 3). Then, the highest SA is beyond the PR Island.

J and PR have approximately the same area. They are macroblocks in the PBZ but their 2 pairs of mesoblocks are differently oriented (J to E-W and PR to N-S). Their MDFO have an E-W main strike but different sinuosity coefficient values (J = 0.69 and PR = 0.59). These data allow considering the same regional tectonic influence but taking into account existing local differences.

## FINAL NOTES AND CONCLUSIONS

In-depth and extensive geomorphic analysis, employing aerial photographs, geomorphologic, geologic, geophysic, topographic and field studies show that the morphology of the Northern Caribbean islands can be linked to lateral variations in the geometry and tectonism of the Caribbean-North American PBZ. The morphotectonic methods applied may develop a homogeneous model where 2 megablocks, 7 macroblocks, 42 mesoblocks, 653 microblocks and 1264 nanoblocks exist, to be achieved. They constitute a set of active blocks under rotation, uplifting and tilting movements. H is the most complex structure. Also, the earthquake occurrence in H is related to the stress concentrations in the vicinity of its different morphotectonic zones.

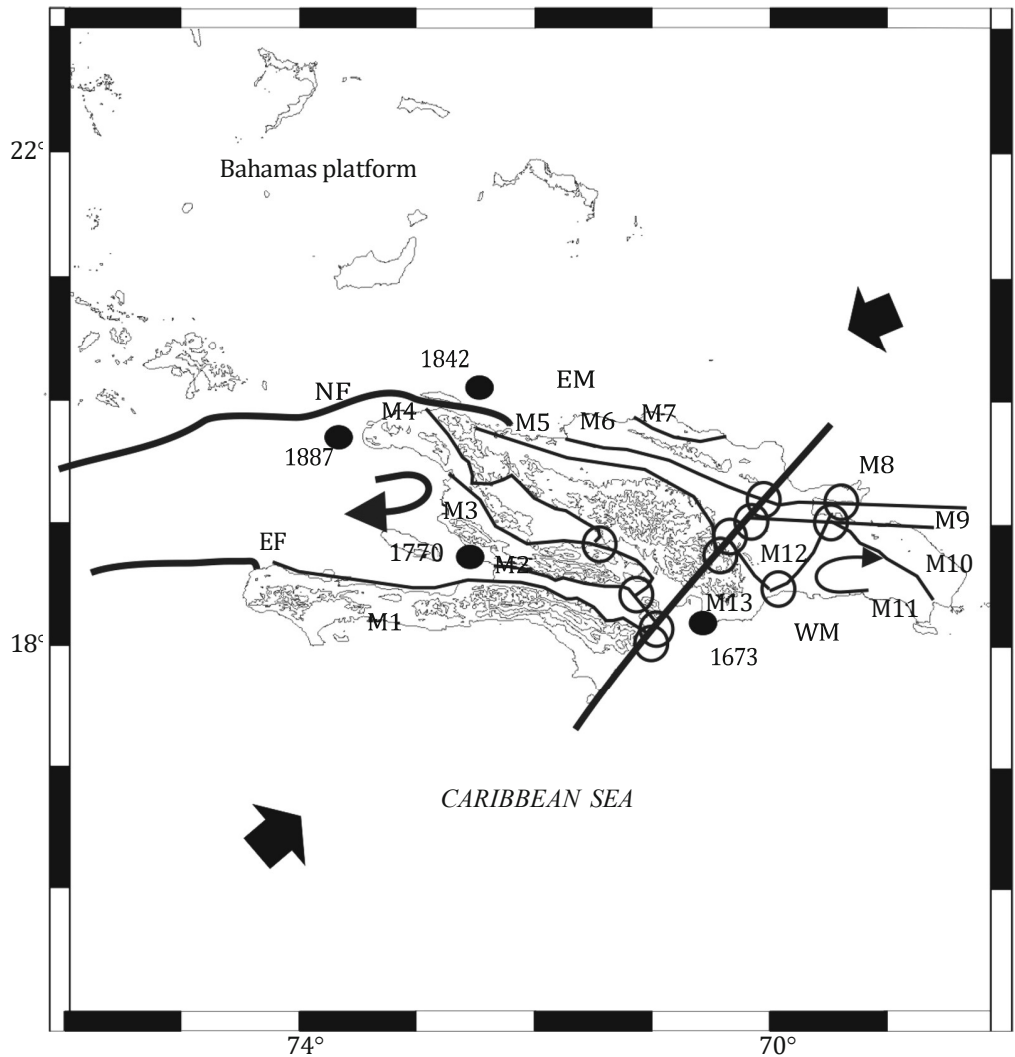

**Fig. 5.** Morphotectonic map of Hispaniola. Black arrow—main sense of movement; black circle—knot; black line—faults—alignments: EF—Enriquillo, NF—Northern; curve black arrow—sense of block movement; macroblocks: EM—Eastern, WM—Western; M1–M13—mesoblock.

In the Table 5 there are 4 parameters that we use to find neotectonic activity in the mentioned islands. Three of these indexes have anomalus values. They are indicated in gray color.

From a neotectonic point of view J, H and PR are structures on the Caribbean plate but C belongs to the North American plate. A total of 13 active knots (N1, N11, NMG0, NMG1, NMG2, NMG3, NMG4, NMG5, NMG6, NMG7, NMC2, NMC3 and NMN1) and 8 cells are the main articulation areas between these structures (Fig. 8). Taking this into account, we made a proposal of 8 geodynamic cells, accommodating the deformations and cutbacks of the crust in this segment of the PBZ, and according very well with the GPS and focal mechanism data. The geodynamic cells are hourly and anti-hourly to display

**Table 5.** Neotectonic indexes

| Parameter | C | H | J | PR | MV |
|---|---|---|---|---|---|
| Fluvial basins | 0.7 | 0.6 | 0.6 | 0.6 | 0.8 |
| Main knots | 0.8 | **1.2** | 0.8 | 0.7 | 0.8 |
| River valleys | 1.4 | 1.7 | **2.2** | 1.1 | 1.5 |
| Uplifting areas | 0.7 | 0.6 | 0.8 | **0.1** | 0.6 |

MV—Middle value.

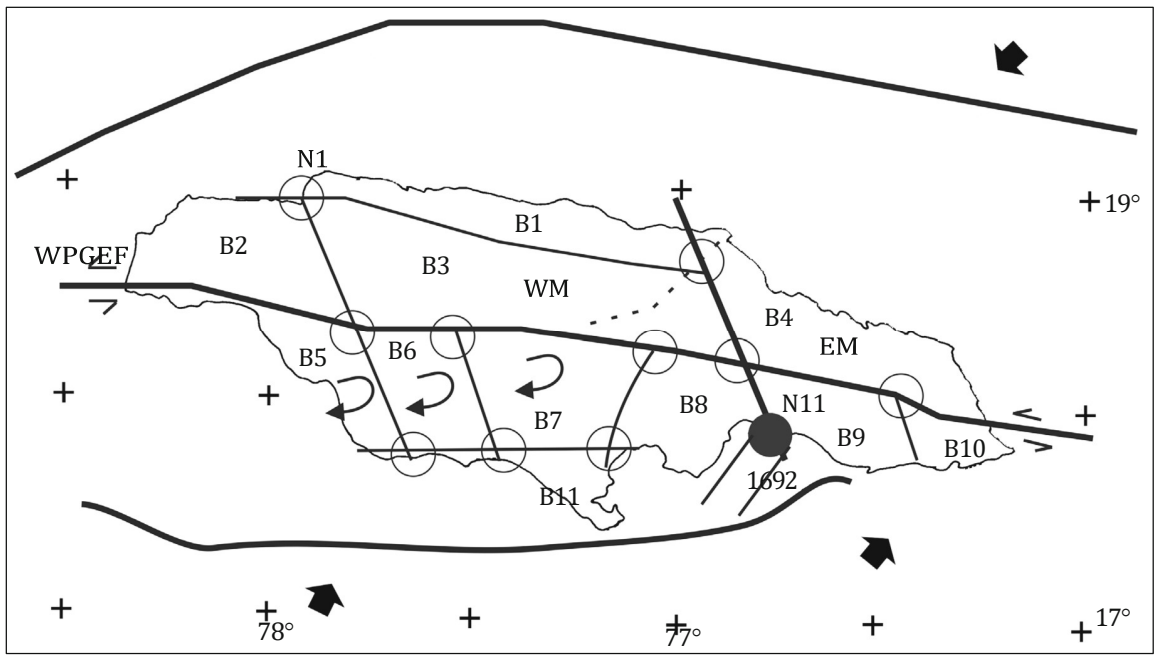

**Fig. 6.** Morphotectonic map of Jamaica. Black arrow is the main direction of movement; black circle—knot; black line—faults—alignments: WPGEF—Walton-Plantain Garden-Enriquillo; curve black arrow—sense of block movement; mesoblocks: EM—Eastern, WM—Western.

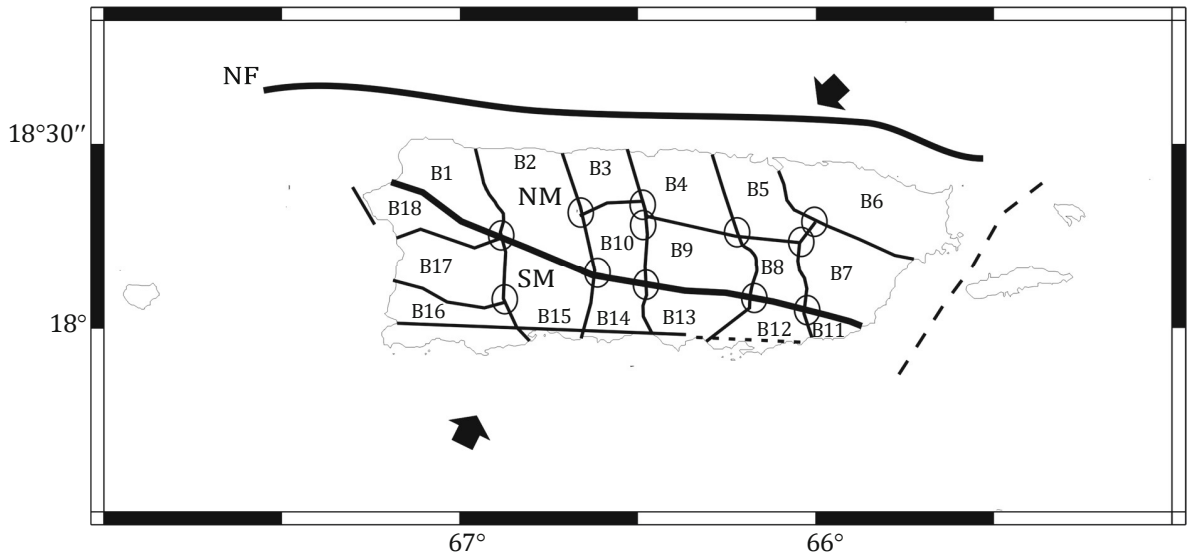

**Fig. 7.** Morphotectonic map of Puerto Rico. Black arrow is the main direction of movement; black circle, knot; black lines, alignments of faults: NF—Northern; curve black arrow, direction of block movement; mesoblocks: NM—Northern, SM—Southern.

a main movement to the NNE. They are composed of the master faults, taking into account the direction of the movement, the majority of them to the left.

There were represented knots of 2 orders in the figures: (1) first (NMG0 – NMG7); (2) second (N1, N11, NMC2, NMC3 and NMC1). We focused on the NMG1 knot because, as mentioned before, it is where 3 large active faults meet. Two of them (Septentrional and Oriente faults) are from the Caribbean plate and the Nortecubana fault belongs to the North American plate. There are some other interesting data as: (1) the highest level of SA (historic and contemporary); (2) the change of fault styles; (3) the relationship to some tsunamis; (4) the relationship with the Bahamas Platform; (5) the clear differences of the gravity values in the Eastern Cuba, H and Oriente trough.

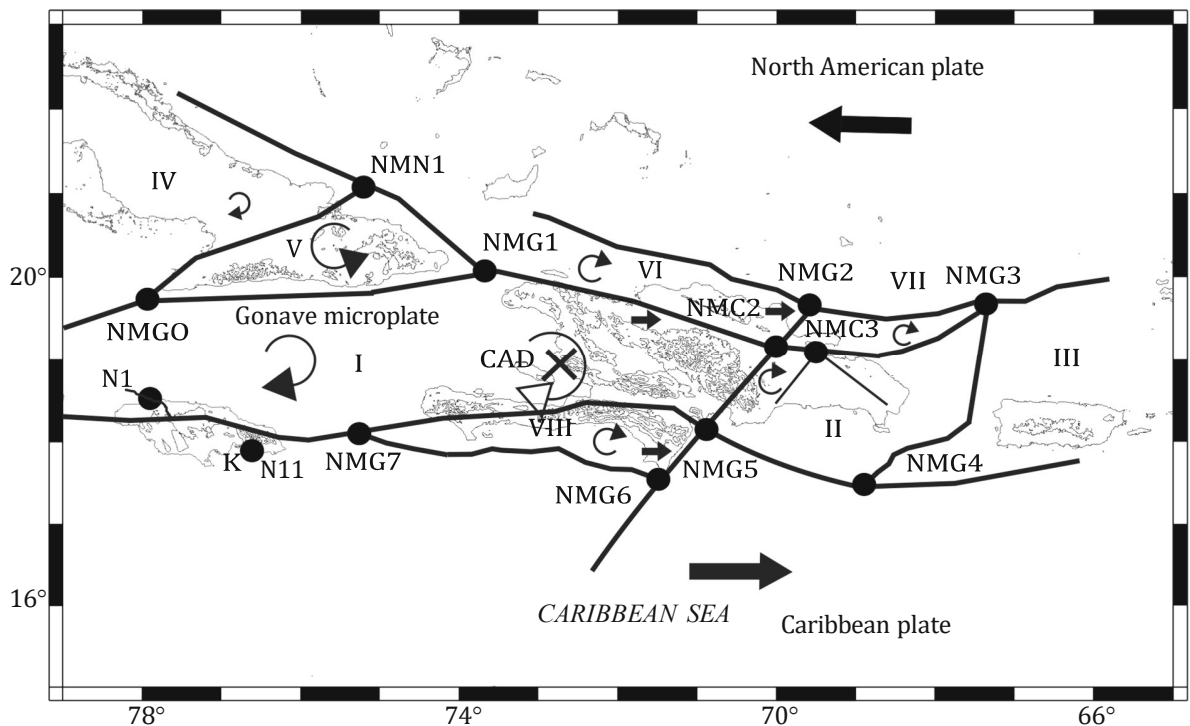

**Fig. 8.** Geodynamic cells in the Northern Caribbean. Black circles denote knots; black curve arrow, direction of cells movement; black lines, alignments of faults; CAD—Central axis of neotectonic deformations; heavy black lines, directions of plate and microplate movements; NMG0, NMG1, NMG2, NMG3, NMG4, NMG5, NMG6, NMG7—knot of first order; NMN1, NMC1, NMC3 are knots of second order; I–VIII are cells.

The NMG0 and NMG1 knots are in the Caribbean plate. They are responsible for the origin of the NMN1 knot in the North American plate, associated with the Nortecubana fault. It produces a dynamic cell in Eastern C where a great number of earthquakes, mainly at 0–20 km of depth, occur.

The majority of the C megablock is tectonically more stable than the rest of the mentioned islands. The major instability area of C is associated with the Eastern macroblock, adjacent to the Oriente fault. SA of the Western and Central-Eastern macroblocks is explained by the stress transmission toward NE from the PBZ. In the northern part of the Central-Eastern macroblock 2 tsunamis were generated (Fig. 4 and Table 3).

The largest seismogenetic structures in the Northern Caribbean are an array of active fault segments. All they are first order disruptive structures. They constitute the two external boundaries of the H, J and PR islands.

The neotectonic is quite different in the Cuban territory. The most important seismogenetic fault (Oriente) is located between the Cabo Cruz and Punta de Maisí, in Eastern macroblock. The northern border (inside of the North American plate) of the Cuban megablock is a large and deformed (>1000 km) 2nd order fault (Nortecubana).

The Nortecubana and Septentrional faults are approximately of the same orientation (NE) and developed in the north shore of C and H, respectively, and related with the Bahamas Platform. They are responsible for tsunamis occurring in both islands. In the S of our study region other faults (Swan-Oriente and Enriquillo-Plantain Garden) are approximately parallel to the formerly mentioned E-W structures. Some strong earthquakes have occurred between the E of C (Punta de Maisí) and the NW of H (Haitian Cape). Three faults meet in this area (Oriente, Nortecubana and Septentrional). They constitute the NMG1 active knot. These two mentioned areas are separated by ~70 km with a similar marine terrace system (lithology and altitude levels).

Exists a direct relationship between strong earthquakes, tsunamis and fatalities to all Caribbean islands, and the maximum value is in H. Roughly 88% of all fatalities produced in the Northern Caribbean by earthquakes, occurred in H. Thus we considered H to be the most active morphotectonic unit.

C and J are emerged structures but on the opposite side of the Oriente trough. From these islands to the trough the marine relief is quite different. The most complex is the Cuban segment. Eastern C and J are the only morphostructures without a subduction zone.

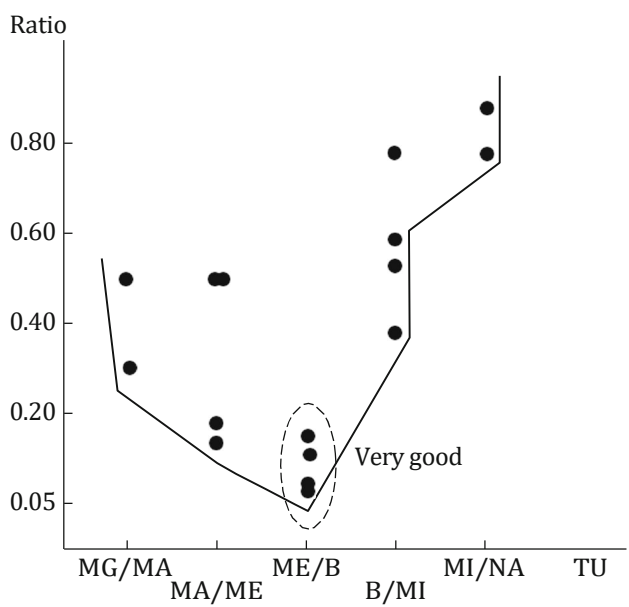

**Fig. 9.** Structural relations. Axis TU—Territorial unit: MG/MA—Megablock/Macroblock; MA/ME—Macroblock/Mesoblock; ME/B—Mesoblock/Block; B/MI—Block/Microblock; MI/NA—Microblock/Nanoblock; Ratio values.

They have had 11 earthquakes, while there were 24 in the H-PR zone. Also, the Mmax is different (Eastern C-J = 7.5 and H-PR = 8.2). J is also related to the SW part of H. They are connected by the active fault system Enriquillo-Plantain Garden.

H and PR are separated by the Mona Passage as an active pull-apart, but connected in the S by the Muertos trough and in the north by the active Septentrional fault zone. PR is characterized by 2 troughs in its northern and southern borders. J has higher altitudinal values, quantity of close relief surfaces and fracture density than PR. Conversely, the number of blocks, fluvial valleys and uplifting areas is higher in PR. J and PR proportionally have the same quantity of superficial TU. C and H show similar characteristic, in addition to a singular anomaly regarding mesoblock/block. J and PR macroblocks have some different characteristics regarding fluvial basins. In J the higher area value is located in the S basin, while in PR it is in the north basin. The MDFO of PR is oriented from SW to NE, but in J the strike is from NW to SE. With these data we confirm the morphotectonic contraposition of the 2 macroblocks in both sides of H.

We have considered the existence of a morphotectonic deformation axis to the Greater Antilles Arc near to Puerto Principe (eastern H). Some strong earthquakes were determined here. It is associated with the Enriquillo fault.

Finally, the most relevant morphotectonic characteristics in the Greater Antilles Arc are: (1) the presence of 2 deep oceanic troughs at the same geographic latitude and separated by ~600 km with different geodynamic conditions but in the same PBZ; (2) the concave figure to the S of C, approximately at the same longitude of the Mid Cayman Spreading Centre; (3) the SW-NE morphostructural escarpment of Beata that limits the Enriquillo fault to the E; (4) the particular location of 2 pairs of structures, at approximately the same distance and latitude, J—PR macroblocks and Oriente—PR troughs, respectively, with respect to the H megablock; (5) the symmetric location of 2 basins (Colombia and Venezuela) in southern H; (6) the heterogeneous geometric figure of H (quite different to the other islands); (7) the highest value of neotectonic uplift in H (2235 m); (8) the fractal dimension values of J and PR are quite similar (~1.8) but less than that of the H = 1.98; (9) the Eastern C macroblock and H megablock have similar neotectonic characteristics; (10) the Western and Central-Eastern C macroblocks are quite different respect to the Eastern one; (11) the superficial rate value (~0.07) in the Caribbean Sea shows a good adjusted; (12) the close relationship between the E border of the Greater Antilles Arc and the subduction zone of the Lesser Antilles Arc; (13) the very good adjustement in the morphotectonic relationship (Fig. 9) between mesoblocks and blocks that reflect a minimum value.

ACKNOWLEDGMENTS

Part of the used funds comes from the TSUJAL (CGL2011-29474-C02-01) and CARIBENORTE (CTM2006-13666-C02-02) projects. Amador García Sarduy drew the figures. The research was carried out in the Departamento de Física de la Tierra, Astronomía y Astrofísica I, Facultad de Ciencias Físicas, Universidad Complutense de Madrid. The German friends Hans Joachim Franzke and Joachim Pilarski stimulated the initial idea of this paper.

REFERENCES

1. V. V. Adushkin, I. A. Sanina, I. S. Vladimirova, Yu. V. Gabsatarov, E. M. Gorbunova, and G. N. Ivanchenko, "Study of neotectonic activity of morphostructures in the central part of East European Craton by remote sensing methods," Izv., Phys. Solid Earth **50**, 169–176 (2014).
2. M. Alekseevskaya, A. Gabrielov, I. Gelfand, A. Gvishiani, and E. Rantsman, "Formal morphostructural zoning in mountain territories," J. Geophys. **43**, 227–233 (1977).
3. M. S. Arnaiz and Y. Garzón, "Nuevos mapas de anomalías gravimétricas del Caribe," Interciencia **37**, 176–182 (2012).
4. B. A. Assinovskaya and S. L. Solovyev, "Definition and description of the sources zone of potential earthquakes in the Barents Sea," Izv., Phys. Solid Earth **29**, 664–675 (1994).
5. W. H. Bakun, C. H. Flores, and U. Ten Brink, "Significant earthquakes on the Enriquillo fault system, His-

paniola 1500–2010: Implication for seismic hazard," Bull. Seismol. Soc. Am. **102**, 18–30 (2012).

6. S. C. Bathia, T. R. K. Chetty, M. Filomonov, A. Gorshkov, E. Rantsman, and M. N. Rao, "Identification of potential areas for the occurrence of strong earthquakes in Himalayan Arc region," Proc. Indian Acad. Sci., Earth Planet. Ser. **101**, 369–385 (1992).
7. K. Burke, C. Cooper, J. F. Dewey, P. Mann, and J. L. Pindell, "Caribbean tectonics and relative plate motions," in *The Caribbean–South American Plate Boundary and Regional Tectonics*, Vol. 162 of *Geol. Soc. Am., Mem.*, Ed. by W. E. Bonini, R. B Hargraves, and R. Shagam (1984), pp. 31–63.
8. D. B. Byrne, G. Suarez, and W. R. McCann, "Muertos trough subduction-microplate tectonics in the northern Caribbean," Nature **317**, 420–421 (1985).
9. E. Calais and B. Mercier de Lepinay, "Strike-slip tectonic processes in the northern Caribbean between Cuba and Hispaniola," Mar. Geophys. Res. **17**, 63–95 (1995).
10. J. E. Case, T. L. Holcombe, and R. G. Martin, "Map of geologic provinces in the Caribbean region," in *The Caribbean–South American Plate Boundary and Regional Tectonics*, Vol. 162 of *Geol. Soc. Am., Mem.*, Ed. by W. E. Bonini, R. B Hargraves, and R. Shagam (1984), pp. 1–31.
11. J. D. Chaytor and U. S. Ten Brink, "Extension in Mona Passage, northeast Caribbean," Tectonophysics **493**, 74–92 (2010).
12. S. Chiesa and G. Mazzoleni, "Dominican Republic (Hispaniola Island, North-Eastern Caribbean): A map of morpho-structural units at a scale 1 : 500000 through Landsat TM image interpretation," Rev. Geol. Am. Centr. **25**, pp. 99–106 (2001).
13. M. O. Cotilla, "Alternative interpretation for the active zones of Cuba," Geotectonics **48**, 459–483 (2014).
14. M. O. Cotilla, "Historia sobre la Sismología del Caribe Septentrional," Rev. Hist. Am. **147**, pp. 111–154 (2012).
15. M. O. Cotilla, *Un recorrido por la Sismología de Cuba* (Editorial Complutense, Madrid, 2007) (in Spanish).
16. M. O. Cotilla and J. L. Álvarez, *Determinación de las zonas de origen de terremotos de La Española y Jamaica* (Instituto de Geofísica y Astronomía, Academia de Ciencias de Cuba, Habana, 1991) (in Spanish).
17. M. Cotilla and D. Córdoba, "Análisis morfotectónico de la Isla de Puerto Rico, Caribe," *Rev. Geofís.* **52**, 79–126 (2011).
18. M. Cotilla and D. Córdoba, "Comments about tsunami occurrences in the northern Caribbean," in *Tsunami—A Growing Disaster*, Ed. by M. Mokhtari (InTech, 2011), Ch. 7.
19. M. O. Cotilla and D. Córdoba, "Determination of lineaments in Hispaniola," Rev. Geogr. **147**, 114–133 (2010).
20. M. O. Cotilla and D. Córdoba, "Study of the Cuban fractures," Geotectonics **44**, 176–202 (2010).
21. M. O. Cotilla and Córdoba D. "Morphostructural analysis of Jamaica," Geotectonics **43**, 420–431 (2009).
22. M. O. Cotilla and D. Córdoba, "The Hispaniola fluvial system and its morphostructural context," Phys. Geogr. **30**, 453–478 (2009).
23. M. Cotilla and A. Udías, Geodinámica del límite Caribe-Norteamérica," Rev. Soc. Geol. Esp. **12**, 175–186 (1999).
24. M. O. Cotilla, D. Córdoba, and M. Calzadilla, "Morphotectonic study of Hispaniola," Geotectonics **41**, 368–391 (2007).
25. M. O. Cotilla, H. J. Franzke, and D. Córdoba, "Seismicity and seismoactive faults of Cuba," Russ. Geol. Geophys. **48**, 505–522 (2007).
26. M. O. Cotilla, P. Bankwitz, J. L. Álvarez, H. J. Franzke, M. F. Rubio, and J. Pilarski, "Cinemática neotectónica de Cuba Oriental," Rev. Soc. Geol. Esp. **11**, 33–42 (1998).
27. M. O. Cotilla, M. Rubio, L. Álvarez, and G. Grünthal, Potenciales sísmicos sector Centro-Occidental del arco de Las Antillas Mayores. Rev. Geofís. **46**, 127–150 (1997).
28. M. O. Cotilla, E. C. González, Cañete C.C., J. L. Díaz, and R. Carral, "La red fluvial de Cuba y su interpretación morfoestructural," Rev. Geogr. **134**, 47–74 (2003).
29. M. Cotilla, P. Bankwitz, H. J. Franzke, L. Álvarez, E. González, J. L. Díaz, G. Grünthal, J. Pilarski, and F. Arteaga, "Mapa sismotectónico de Cuba, escala 1 : 1000000," Comun. Cient. Geofís. Astron., no. 23 (1991).
30. M. O. Cotilla, E. C. González, H. J. Franzke, J. L. Díaz, F. Arteaga, and L. Álvarez, "Mapa neotectónico de Cuba, escala 1 : 1000000," Comun. Cient. Geofís. Astron., no. 22 (1991).
31. C. De Mets, R. G. Gordon, and D. F. Argus, "Geologically current plate motions," Geophys. J. Int. **181**, 1–80 (2010).
32. C. De Mets, P. E. Jansma, G. S. Mattioli, T. Dixon, P. Farina, R. Bilham, E. Calais, and P. Mann, "GPS geodetic constraints on Caribbean-North American plate motion," Geophys. Res. Lett. **27**, 437–440 (2000).
33. W. P. Dillon, J. A. Austin Jr., K. M. Scanlon, N. T. Edgar, and L. M. Parson, "Accretionary margin of the north-western Hispaniola: Morphology, structure, and development of the northern Caribbean plate boundary," Mar. Pet. Geol. **9**, 70–88 (1992).
34. T. H. Dixon, F. Faina, F. De Mets, P. Mann, and E. Calais, "Relative motion between the Caribbean and North American plates and related boundary deformation from a decade of GPS observation," J. Geophys. Res.: Solid Earth **103**, 15157–15182 (1998).
35. J. F. Dolan and D. J. Wald, "The 1943–1953 north-central Caribbean earthquakes: Active tectonic setting, seismic hazard and implications for Caribbean-North America plate motion," in *Active Strike-Slip and Collisional Tectonics of the Northern Caribbean Plate Boundary Zone*, Vol. 326 of *Geol. Soc. Am., Spec. Pap.*, Ed. by J. F. Dolan and P. Mann (1998), pp. 143–170.
36. J. F. Draper, P. Mann, and J. W. Lewis "Hispaniola," in *Active Strike-Slip and Collisional Tectonics of the Northern Caribbean Plate Boundary Zone*, Vol. 326 of Geol. Soc. Am., Spec. Pap., Ed. by J. F. Dolan and P. Mann (1998), pp. 129–150.
37. N. W. Driscoll and J. B. Diebold, "Deformation of the Caribbean region: One plate or two?," Geology **26**, 1043–1046 (1998).
38. T. W. Donnelly, "The Caribbean sea floor," in *Caribbean Geology: An Introduction*, Ed. by S. K. Donovan and T.A. Jackson (UWI Publ. Assoc., Kingston, 1994), pp. 41–64.

39. N. V. Dumistrashko and D. A. Lilienberg, *Geomorphic Methods in Seismotectonic Research* (Akad. Nauk SSSR, Moscow, 1954) (in Russian).
40. G. Giunta, L. Beccaluva, and F. Siena, "Caribbean plate margin evolution: Constraints and current problems," Geol. Acta **4**, 265–277 (2006).
41. E. González, M. Cotilla, C. Cañete, R. Carral, J. Díaz and F. Arteaga, "Estudio morfoestructural de Cuba," Geogr. Fis. Din. Quat. **26**, 49–69 (2003).
42. G. P. Gorshkov, *Regional Seismotectonics of the Southern USSR* (Nauka, Moscow, 1984) (in Russian).
43. A. I. Gorshkov, I. V. Kuznetsov, G. F. Panza, and A. A. Soloviev, "Identification of future earthquake sources in the Carpatho-Balkan orogenic belt using morphostructural criteria," Pure Appl. Geophys. **157**, 79–95 (2000).
44. I. E. Gubin, *Regularities of Seismic Manifestations in the Territory of Tajikistan* (Izd. Akad. Nauk SSSR, Moscow, 1960) (in Russian).
45. A. D. Gvishiani, B. A. Dzeboev, and S. M. Agayan, "A new approach to recognition of the strong earthquake-prone areas in the Caucasus," Izv., Phys. Solid Earth **49**, 747–766 (2013).
46. Instituto Sismológico Universitario, República Dominicana. (uasd.edu.do/index/php/es/)
47. International Seismological Centre. http://www.isc.ac.uk/.
48. P. E. Jansma, G. S. Mattioli, A. López, C. De Mets, T. H. Dixon, P. Mann, and E. Calais, "Neotectonics of Puerto Rico and the Virgin Islands, northeastern Caribbean, from GPS geodesy," Tectonics **19**, 1021–1037 (2000).
49. P. Mann and K. Burke, "Neotectonics of the Caribbean," Rev. Geophys. Space Phys. **22**, 309–392 (1984).
50. P. Mann, F. W. Taylor, E. Lawrence, and K. U. Teh-Lung, "Actively evolving microplate formation by oblique collision and sideways motion along strike-slip faults. An example from the northeastern Caribbean plate margin," Tectonophysics **246**, 1–69 (1995).
51. W. R. McCann, "The Muertos trough as a major earthquake and tsunami hazard for Puerto Rico," EOS, Trans., Am. Geophys. Union **88** (23), Jt. Assembly Suppl. Abstr. S52A-03 (2007).
52. W. R. McCann and W. D. Pennington, "Seismicity, large earthquakes, and the margin of the Caribbean plate," in *The Geology of North America*, Vol. H: *The Caribbean Region*, Ed. by G. Dengo and J. E. Case (Geol. Soc. Am., Boulder, CO, 1990), Ch. 4, pp. 291–306.
53. P. Molnar and L. R. Sykes, "Tectonics of the Caribbean and Middle America regions from focal mechanisms and seismicity," Geol. Soc. Am. Bull. **80**, 1639–1684 (1969).
54. S. P. Nishenko, W. R. McCann, and M. D. Wiggins-Grandison, "An active restrainning bend on the eastern extension of the Swan Fracture Zone," Geol. Soc. Am. Abstr. Progr. **27**, (1996).
55. D. Núñez, D. Córdoba, M. O. Cotilla, and A. Pazos, "Modeling the crust and upper mantle in northern Beata Ridge (CARIBE NORTE Project)," Pure Appl. Geophys. **173**, 1–23 (2016).
56. P. G. Okubo and K. Aki, "Fractal geometry in the San Andreas fault system," J. Geophys. Res., B **92**, 345–355 (1987).
57. M. Pubellier, A. Mauffret, S. Leroy, J. M. Vila and H. Amilcar, "Plate boundary readjustment in oblique convergence: Example of the Neogene of Hispaniola, Greater Antilles," Tectonics **19**, 630–648 (2000).
58. C. S. Prentice, P. Mann, A. J. Crone, R. D. Golde, K. W. Hudmut, R. W. Briggs, R. D. Koeler, and P. Jean, "Seismic hazard of the Enriquillo-Plantain Garden fault in Haiti inferred from paleoseismology," Nat. Geosci. **3**, 789–793 (2010). doi 10.1038/NE0991
59. Puerto Rico Seismic Network. http://redsismica.uprm.edu/.
60. E. Ya. Rantsman, *Earthquake Locations and Morphostructures of Mountain Areas* (Nauka, Moscow, 1979) (in Russian).
61. C. M. Roig-Silva, E. Asencio, and J. Joyce, "The northwest trending North Boquerón Bay-Punta Montalva fault zone: A trough going active fault system in southwestern Puerto Rico," Seismol. Res. Lett. **84**, 538–550 (2013).
62. M. Rubio, M. Cotilla, and L. Álvarez, *Evidencias sobre la microplaca Gonave. Informe científico-técnico* (Instituto de Geofísica y Astronomía, Academia de Ciencias de Cuba, Habana, 1994) (in Spanish).
63. E. Ruellan, B. Mercier de Lépinay, M. O. Beslier, M. Sosson, C. Monnier, S. Leroy, D. Rove, and G. Cruz Calderón, "Morphology and tectonics of the Mid-Cayman spreading centre," in *Abstracts of the Joint EGS-AGU-EUG Assembly, Nice, France, 2003*.
64. Y. K. Shchukin and T. E. Lyustikh, *Geodynamics and Seismicity. General Geology*, Vol. 14 of *Results in Science and Technology* (VINITI, Moscow, 1981) (in Russian).
65. L. R. Sykes, W. R. McCann, and A. L. Kafka, "Motion of Caribbean plate during last 7 million yr and implications for earlier Cenozoic movements," J. Geophys. Res., B **87**, 10656–10676 (1982).
66. S. Taber, "The seismic belt in the GReater Antilles," Bull. Seismol. Soc. Am. **12**, 199–219 (1922).
67. US Geological Survey (USGS). www.usgs.gov.
68. S. A. Ushakov, A. I. Avgaev, Yu. I. Galushkin and E. P. Dubikin, "Isostasy destruction of the Caribbean lithosphere and the geodynamic analysis," in *Tectonics and Geodynamics of the Caribbean Region* (Nauka, Moscow, 1979), pp. 63–77.
69. S. R. Van Dusen and D. I. Doser, "Faulting process of historic (1917–1962) M ≥ 6.0 earthquake along the North-Central Caribbean margin," Pure Appl. Geophys. **157**, 719–736 (2000).
70. L. P. Vinnik, A. A. Lukk, and I. L. Nersesov, "Nature of the intermediate seismic zone in the mantle of Pamirs–Hundu-Kush," Tectonophysics **38**, 35–47 (1977).
71. G. K. Westbrook, H. P. Boot and J. H. Peacock, "Lesser Antilles subduction zone in the vicinity of Barbados," Nat. Phys. Sci. **244**, 118–120 (1973).
72. Wice D., Funcilio R., Porotto M. and Salvani F. "Features and lineament orientations in Haly," Geol. Soc. Am. Bull. **96**, 112–138 (1985).
73. M. Wiggins-Grandison and K. Atakan, "Seismotectonics of Jamaica," Geophys. J. Int. **160**, 573–580 (2005).
74. M. P. Zhidkov, I. M. Rotvain, and A. M. Sadowskii, "Forecast of more probable sites for earthquake occurrences. Multiple interceptions of lineaments in the Armenian area," Vychisl. Seismol., No. 8, pp. 53–70 (1975).